\documentclass{aa}  
\usepackage{enumerate}
\usepackage{upgreek}

\usepackage{stackengine}
\usepackage{scalerel}
\usepackage{color,psfrag}
\usepackage{graphicx}
\usepackage{tkz-euclide}
\usepackage{tikz}
\usetikzlibrary{calc}
\usetikzlibrary{decorations.pathmorphing}
\usetikzlibrary{shapes.geometric}
\usetikzlibrary{matrix}
\usetikzlibrary{arrows,decorations.markings}
\usetikzlibrary{shadings,intersections}
\usepackage{natbib}
\bibpunct{(}{)}{;}{a}{}{,} 
\usepackage{txfonts}
\usepackage{hyperref}
\usepackage{lipsum}
\usepackage{subcaption}         
                                
\usepackage{lscape}             
                                
\usepackage{placeins}

\begin{document}

   \title{Gravitational Lensing Signatures of Hayward-like Black Holes}

   \author{Chen-Hung Hsiao\inst{1}\corrauth{chsiao@fudan.edu.cn}        
        \and Limei Yuan\inst{1}\email{lmyuan24@m.fudan.edu.cn}
        \and Yidun Wan\inst{1,2}\corrauth{ydwan@fudan.edu.cn}
        }

   \institute{State Key Laboratory of Surface Physics, Center for Astronomy and Astrophysics, Department of Physics, Center for Field Theory and Particle Physics, and Institute for Nanoelectronic devices and Quantum computing, Fudan University, Shanghai 200433, China\and Hefei National Laboratory, Hefei 230088, China}

   \date{\today}

  \abstract
    {  The Hayward-like black hole~\cite{Calza:2025mrt}, a nonsingular metric that resolves the Cauchy horizon issue, presents a potentially viable alternative to the classical Schwarzschild solution.}
   { We aim to examine whether the Hayward-like nonsingular black hole produces distinct gravitational lensing signatures in both weak and strong-field regimes that could distinguish it from a Schwarzschild black hole using current or future astronomical facilities.
   } 
   {We utilize the Gauss-Bonnet theorem method \cite{Gibbons:2008rj} to analyze weak-field deflection and the Strong Deflection Limit (SDL) method~\cite{Bozza:2001xd,Bozza:2002zj} to compute SDL observables ($\theta_{\infty}, s, r_{\mathrm{mag}}$) and time delays ($\Delta T_{2,1}$).}
   {In the weak-field limit, the deflection angle includes a small positive correction proportional to $m\ell^{2}/b^{3}$, indicating slightly stronger light bending than in Schwarzschild, though the effect remains observationally negligible at large impact parameters. Current galaxy-scale Einstein-ring data, such as from ESO~325-G004, cannot yet constrain the regular-core scale $\ell$. In the strong-deflection regime, for Sgr~A* and M87*, the asymptotic position $\theta_{\infty}$ is identical to Schwarzschild’s. Nevertheless, $\ell$ modifies strong-lensing coefficients $\bar a, \bar b$, influencing angular separations $s$, relative flux ratio $r_{\mathrm{mag}}$, and time delays $\Delta T_{2,1}$. Our predicted values for these observables remain consistent with current data, suggesting that future high-precision measurements of strong-field lensing may distinguish Hayward-like from Schwarzschild black holes.}
  {}

   \maketitle
\nolinenumbers

\section{Introduction}
Gravitational lensing serves as a powerful probe of compact objects, sensitive to the spacetime geometry in both the weak- and strong-field regimes. The Hayward-like black hole, a nonsingular metric that resolves the Cauchy horizon issue, presents a potentially viable alternative to the classical Schwarzschild solution. A key question is whether its distinct geometry yields observable lensing signatures that could distinguish it from a Schwarzschild black hole with current or future astronomical facilities. In this paper, we systematically investigate the gravitational lensing by a Hayward-like black hole across both weak- and strong-field regimes to address this question. We obtain the following key results:
\begin{enumerate}
    \item  In the weak-field region, the deflection angle increases with the regular-core scale $\ell$. This represents a positive deviation from the standard Schwarzschild lensing scenario, where $\ell=0$. The Analysis of the SDL reveals parameters for computing the analytical deflection angles in the strong-field regime. 
    
    \item  The SDL observables ($\theta_{\infty}, s, r_{\mathrm{mag}}$) are computed using data corresponding to M87* and Sgr~A*. For both black hole candidates, the angular separation $s$ and the relative flux ratio $r_{\mathrm{mag}}$ are $\ell$-dependent. Notably, the asymptotic position $\theta_{\infty}$ was found to be independent of the $\ell$. 
    
    \item The calculated shadow size of Hayward-like black hole remains consistent with the EHT measurements for M87* and Sgr A*. The time delay between the first and second relativistic images, $\Delta T_{2,1}$, increases monotonically with $\ell$.
\end{enumerate}

\textbf{Background and motivation}
Gravitational lensing (GL) is an important tool for studying astrophysical phenomena and testing gravitational theories \cite{Wambsganss:1998gg,ORBi-f3f7ce61-7628-4d1a-b937-387c26218fd7, Prat:2025ucy,Bartelmann:2016dvf}.  When light travels far away from the massive object (lens), the lensing takes place at the weak field region; however, when light propagates close to the gravitational object---where the gravitational field is so prominent that the angular deflection becomes divergent at a particular limit---it is known as the strong deflection limit \cite{Bozza_2002,Bozza:2001xd,Bozza:2002zj,Bozza:2003cp,Bozza:2008ev}.

\begin{figure}[ht!]
\centering
\begin{tikzpicture}[>=latex, scale=0.85,
    image/.style={very thick, blue},
    source/.style={thick, red, densely dashed},
    axis/.style={gray, dashed}]

    \coordinate (O) at (0,0);       
    \coordinate (L) at (0,6);       
    \coordinate (S) at (1.22,11.5);   
    \coordinate (I) at (4.5,9.9);    
    \coordinate (Q) at (3.4, 1);
    \coordinate (J) at (2.0, 6.045);  
    \coordinate (Y) at (2.20, 4.83);  

    \draw[axis] (0,0) -- (0,12) node[above]{optical axis}; 
    \node[right, gray] at (2.3,11.5) {source plane};
    \draw[axis] (-1.2, 11.5) -- (2.2, 11.5); 

    \draw[thick] (L) circle (0.16);
    \fill (L) circle (1pt) node[left=4pt] {$L$};

    \draw[image] (O) .. controls (2.4, 5.35) and (2.4, 6.35) .. (S);
    \draw[source] (O) -- (S);

    \draw[dashed, blue, thick] (S) -- (Q) node[below, black] {$Q$};
    \draw[dashed, blue, thick] (O) -- (I) node[above, black] {$I$};

    \draw[blue,dashed,very thick] (L) -- node[above] {$\rho_0$} (J); 
    \draw[blue,dashed,very thick] (L) -- node[below] {$b$} (Y);  
    \node[right] at (Y) {$Y$};

    \node at (2.7, 7.4) {$\alpha$};
    \draw[black] (2.2, 6.8) arc[start angle=115, end angle=65, radius=1.0];

    \draw[blue] (0, 0.9) arc[start angle=90, end angle=73, radius=0.9];
    \node[blue] at (0.3, 1.1) {$\theta$};
    
    \draw[red] (0, 1.5) arc[start angle=90, end angle=84.5, radius=1.5];
    \node[red] at (0.15, 1.8) {$\beta$};

    \draw[<->] (-0.7,0) -- node[right] {$D_L$} (-0.7,6);
    \draw[<->] (-0.7,6) -- node[right] {$D_{LS}$} (-0.7,11.5);
    \draw[<->] (-1.2,0) -- node[left] {$D_S$} (-1.2,11.5);

    \fill (O) circle (1.5pt) node[below] {$O$};
    \fill (I) circle (1.5pt);
     \fill (Q) circle (1.5pt);
    \fill (S) circle (1.5pt) node[right] {$S$};

\end{tikzpicture}
\caption{Gravitational lensing geometry, presented in the equatorial ($x$-$y$) plane of the lens: $O$, $L$, and $S$ denote the positions of the observer, lens, and source, respectively. $D_L$, $D_S$, and $D_{LS}$ are the angular-diameter distances. $SQ$ and $OI$ represent the tangents to the null geodesic at the source and observer positions, respectively. $LY$ is the perpendicular from the lens $L$ to the tangent $OI$, representing the impact parameter $b$. $\alpha$ is the deflection angle.}
\label{fig:lens_geometry}
\end{figure}
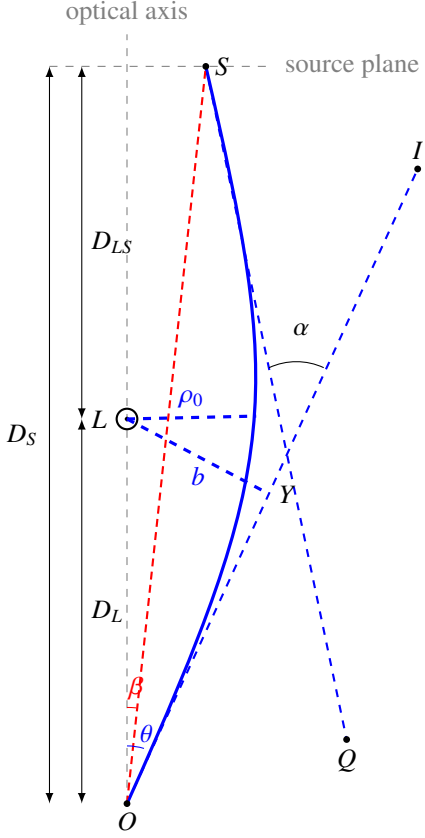

 Figure~\ref {fig:lens_geometry} depicts gravitational lensing geometry, where the reference (optical) axis is one joining the observer $O$ and the lens $L$. The spacetime under consideration, with the lens causing strong curvature, is asymptotically flat; the observer and the source are situated in the flat spacetime region. A light ray emitted from the source $S$ at the angular position $\beta$ is deflected by the lens $L$ and observed by $O$ at the angular position $\theta$. The amount of deflection is measured by the deflection angle $\alpha$.

Black holes are among the most important gravitational lenses.
The deflection angle due to a black hole can be obtained directly from the metric of the black hole. 
 For simplicity, we consider a lens described by a static spherically symmetric metric:
 \begin{equation}
 ds^2 = -B(\rho)dt^{2} + A(\rho)d\rho^{2} + D(\rho)\rho^{2}\left(d\theta^{2}+\sin^{2}\theta d\varphi^{2}\right)\;.
 \label{eq:metric_lens}
 \end{equation}
Since the metric of the black hole is static and spherically symmetric, one can consider the null trajectories to be confined to the equatorial plane. 
Given the metric, the deflection angle can be obtained from the null geodesic equation as \cite{Virbhadra:1998dy,Weinberg:1972kfs}
\begin{equation}
\alpha = 2\Delta\varphi(\rho_0) - \pi\;,
\end{equation}
where the factor of 2 is a consequence of the symmetry of the setup for the observer and the source (the trajectory from the source to the lens is symmetric to that from the lens to the observer), and $\Delta\varphi(\rho_0)$ is the change in azimuthal angle from the closest coordinate to infinity expressed as \cite{Virbhadra:1998dy,Weinberg:1972kfs}

 \begin{equation}
 \Delta\varphi(\rho_0) = \int^{\infty}_{\rho_0} \frac{\left(\frac{A(\rho)}{D(\rho)}\right)^{1/2}}{\left(\frac{D(\rho)B(\rho_{0})}{D(\rho_{0})B(\rho)}\frac{\rho^{2}}{\rho_{0}^{2}}-1\right)^{1/2}}\frac{d\rho}{\rho}\;.
 \label{eq:deflection_angle_generic}
\end{equation}
The black hole's gravitational field deflects a light ray and causes a change in the cross-section of a bundle of rays. The magnification of an image is defined as the ratio of the flux of the image to that of the unlensed source. According to Liouville’s theorem, the surface brightness is preserved as the light is lensed by a black hole. Thus, the magnification of an image is the ratio of the solid angles of the image and the unlensed source:
\begin{equation}
\mu := \frac{d\Omega_{\theta}}{d\Omega_{\beta}} = \left(\frac{\sin\beta}{\sin\theta}\frac{d\beta}{d\theta}\right)^{-1}\;.
\label{eq:magnification_general}
\end{equation}

The Ohanian lens equation, which relates the distances between the observer, lens, and source planes, and its variations are the most reliable exact geometric relations used in strong lensing, as discussed by Bozza~\cite{Bozza:2008ev}. This equation reads
\begin{equation}
D_S \tan\beta = \frac{D_L \sin \theta - D_{LS}\sin(\alpha-\theta)}{\cos(\alpha-\theta) }\;,
\label{eq:lens_equation}
\end{equation}
where $D_L$ is the angular-diameter distance from the observer to the lens, and $D_{LS}$ is the distance from the lens to the source plane, with $D_S=D_L+D_{LS}$.

Einstein’s general relativity (GR) breaks down at the singularity of a black hole. Historically, non-singular black holes, such as the Bardeen and Hayward black holes \cite{Hayward:2005gi,1968qtr..conf...87B}, have been studied. 
 The gravitational lensing effect of non-singular black holes has been studied \cite{Eiroa:2010wm,Zhao:2017cwk,Liu:2026yyj,10.1088/1674-1056/ae29fa}.  
 A recent work~\cite{Calza:2025mrt} constructs Hayward-like non-singular black holes that resolve the Cauchy horizon issue in the original Hayward black hole \cite{Poisson:1990eh}. A Hayward-like black hole has the metric
\begin{equation}
ds^2=-\left(1-\frac{2M(\rho)}{\rho}\right)dt^{2}
+\left(1-\frac{2M(\rho)}{\rho}\right)^{-1}d\rho^{2}
+r^2(\rho)\,d\Omega^{2}\,,
\label{eq:Hayward_like_metric}
\end{equation}
where
\begin{equation}
M(\rho)=\frac{m\,\rho^{3}}{\rho^{3}+2m\ell^{2}},
\qquad
r(\rho)
=\rho+\frac{2m\ell^{2}}{\rho^{2}}\,.
\end{equation}
and satisfies the condition that $\rho \geq \rho_0$, with $\frac{\mathrm{d}r(\rho)} {\mathrm{d}\rho}|_{\rho_0}=0$.
 Here, $\ell$ is the regular-core (length) scale and has an upper bound $\ell < \frac{4m}{3\sqrt{3}}$, and $m$ is the ADM mass parameter (in geometric units). Setting $\ell=0$ reduces~\eqref{eq:Hayward_like_metric} to the Schwarzschild metric. More details of a Hayward-like black hole are given in Appendix \ref{sec:topological_BH}.

In what follows, we shall take the Hayward-like black holes as lenses to study their gravitational lensing effects, in order to discern them from the Schwartzschild black holes using the near-future observational data.

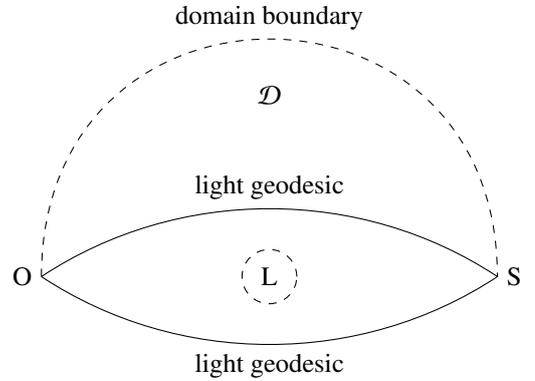
\begin{figure}[!ht]
\centering
     \begin{tikzpicture}[>=Latex,scale=1.2,
    image/.style={very thick,blue},
    source/.style={thick,red,densely dashed},
    axis/.style={gray,dashed}]

  \coordinate (O) at (-2.5,0);
  \coordinate (S) at (2.5,0);
  \coordinate (L) at (0,0);

  \draw (-2.5,0) .. controls (-1,1) and (1,1) .. (2.5,0) node[midway, above] {light geodesic};
  \draw (-2.5,0) .. controls (-1,-1) and (1,-1) .. (2.5,0) node[midway, below] {light geodesic};

  \draw[dashed] (-2.5,0) .. controls (-2.5,3.5) and (2.5,3.5) .. (2.5,0) node[midway, above] {domain boundary};

  \node[right] at (S) {S};
  \node at (L) {L};
  \node[left] at (O) {O};

  \draw[dashed] (L) circle (0.3);

  \node at (0,2.0) {$\mathcal{D}$};

\end{tikzpicture}

	\caption{Lensing geometry in the GBT method. Boundary of domain $D$ lies in a region asymptotically flat with respect to the lens.}
	\label{fig:GBT_graph}
\end{figure}
\section{Weak-field approximation by GBT and Einstein Ring}
\label{sec:weak_GBT}

 We first consider the weak-field scenario. Far from the lens -- black hole -- is regarded as the \emph{weak field} region, where the lens' gravitational field is weak. When light is deflected in this region, the deflection angle $\alpha$ can be expanded as a Taylor series in $1/r$ or, equivalently, in $1/b$, where $b$ is the impact parameter. A powerful approach to compute the weak-field deflection is the Gauss-Bonnet theorem (GBT) method \cite{Gibbons:2008rj}. Appendix \ref{appendix:GBT} briefly reviews the GBT method.
Figure \ref{fig:GBT_graph} depicts the lensing geometry for the GBT method.

\begin{figure}[h!]
\centering
\includegraphics[width=0.45\textwidth]{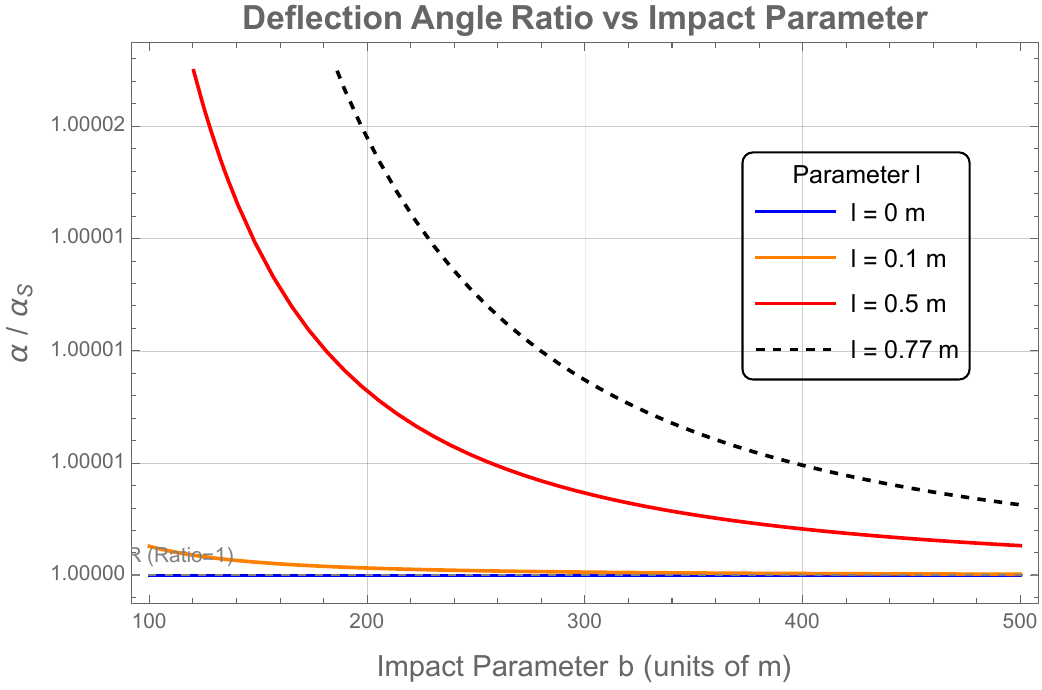}

	\caption{Deflection angle ratio between Hayward-like black hole and Schwarzschild black hole plotted as a function of $b$ for different values of $\ell=0.1m, 0.5m, 0.77m$.} 
	\label{fig:deflection_ratio}
\end{figure}

\begin{figure}[h!]
\centering

\includegraphics[width=0.45\textwidth]{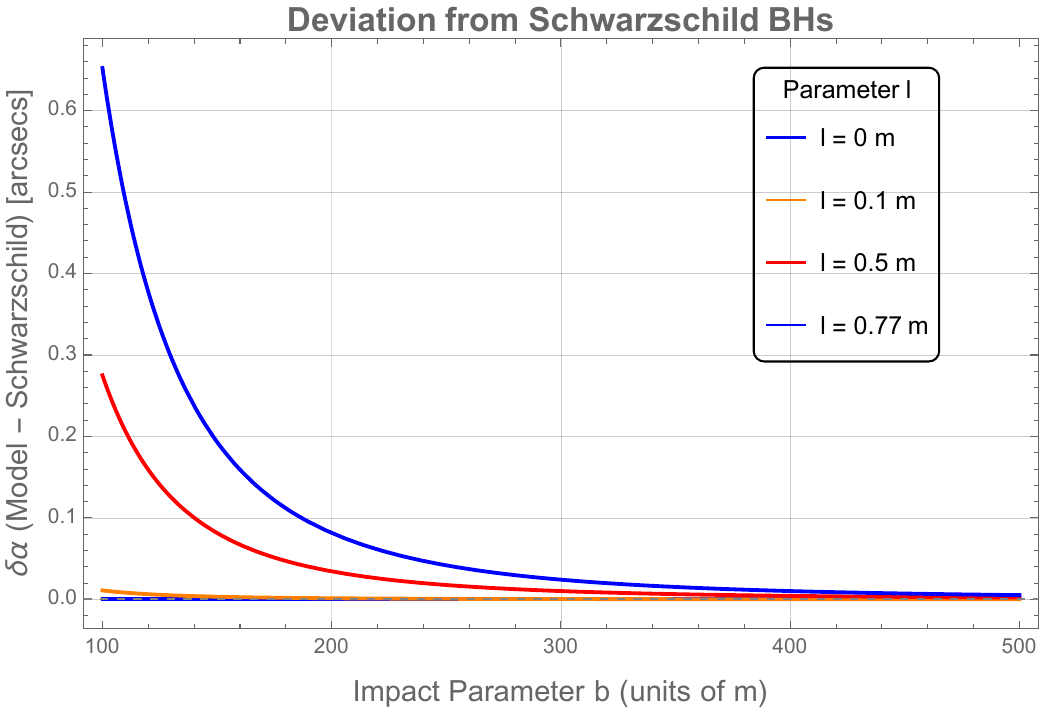}
  
	\caption{Difference of deflection angle between Hayward-like black hole and Schwarzschild black hole is positive, showing positive deviation from the Schwarzschild case.} 
	\label{fig:deflection_Hayward}
\end{figure}

\begin{figure}[h!]
\centering
\includegraphics[width=0.45\textwidth]{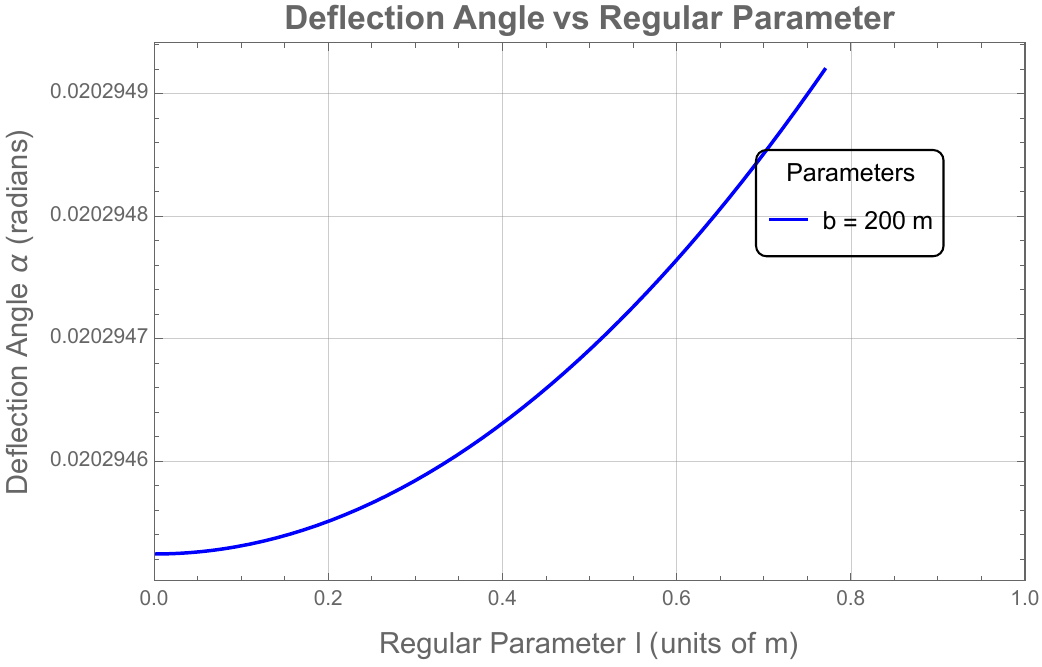}

	\caption{ Deflection angles in the weak-field region as a function of $\ell$. We set $b=200$ m (geometric mass).} 
    \label{fig:deflection_impact_ell} 
\end{figure}

The deflection angle can be obtained by integrating the Gaussian curvature $K$ of region $D$ in Fig. \ref{fig:GBT_graph}:
\begin{equation}
\alpha=-\iint_{D}K\,dS
=-\int_0^{\pi}d\varphi\int_{\rho_{\min}}^{\infty}K(\rho)\sqrt{h}\,d\rho\,,
\label{eq:deflection_GBT}
\end{equation}
For the Hayward-like metric \eqref{eq:Hayward_like_metric}, $K$ and the determinant of the induced metric $h$ $\sqrt{h}$ are expressed as 
\begin{equation}
\begin{split}
K(\rho)&=-\frac{2m}{\rho^{3}}+\frac{3m^{2}}{\rho^{4}}-\frac{12m\ell^{2}}{\rho^{5}}
+\frac{88m^{2}\ell^{2}}{\rho^{6}}\\
&-\frac{120m^{3}\ell^{2}}{\rho^{7}}
+\frac{24m^{2}\ell^{4}}{\rho^{8}}+\mathcal{O}\!\left(\rho^{-9}\right),\\
\sqrt{h}&=\rho+3m+\frac{15m^{2}}{2}\frac{1}{\rho}
+\left(2m\ell^{2}+\frac{35m^{3}}{2}\right)\frac{1}{\rho^{2}}
+\mathcal{O}\!\left(\rho^{-3}\right),
\end{split}
\end{equation}
which reduces to those in the Schwarzschild case at $\ell=0$.
At leading order, the light trajectory may be approximated by a straight line, giving
\begin{equation}
u_{\min}\equiv \frac{1}{\rho_{\min}}\equiv \frac{\sin\varphi}{b}\,,
\end{equation}
where $\varphi$ is the azimuthal angle in the metric and $\rho_{\min}$ is the minima radial coordinate correspond to given impact parameter $b$. To include higher-order corrections, one must expand $u_{\min}$ in powers of $1/b$~\cite{Ono:2017pie}. From the null geodesic equations we find that
\begin{equation}
u_{\min}(\varphi)
=\frac{\sin\varphi}{b}
+\frac{m}{b^{2}}\bigl(1+\cos^{2}\varphi\bigr)
+\frac{4m\ell^{2}}{b^{4}}\bigl(1+\cos^{2}\varphi\bigr)\,.
\end{equation}
Changing the integration variable from $\rho$ to $u=1/\rho$, Eq. \eqref{eq:deflection_GBT} becomes
\begin{equation}
\alpha
=-\int_{0}^{\pi}d\varphi\int_{0}^{u_{\min}(\varphi)}K(u)\sqrt{h}\,u^{-2}\,du\,,
\end{equation}
and expanding consistently in $1/b$, we obtain that to the first correction in $m^2$ and $m\ell^2$,
\begin{equation}
\alpha
\simeq \frac{4m}{b}
+\frac{15\pi}{4}\frac{m^{2}}{b^{2}}
+\frac{16}{3}\frac{m\ell^{2}}{b^{3}}
+\mathcal{O}\!\left(m^3,m^2\ell^2\right).
\label{eq:deflection_angle}
\end{equation}
The leading term $\alpha_{\rm Schw}\simeq 4m/b$ is the standard weak-field Schwarzschild result and depends only on the ADM mass parameter. The next term, $\frac{15\pi}{4}\frac{m^2}{b^2}$, is the usual post-Newtonian correction for a Schwarzschild lens. In contrast, the first imprint of the regular-core scale $\ell$ appears at order $b^{-3}$, $\Delta\alpha_\ell\sim m\ell^2/b^3$, and has therefore a small but positive contribution in the weak-field regime.  Figures~\ref{fig:deflection_ratio} and \ref{fig:deflection_Hayward} show this behavior by comparing $\alpha(b)$ for the Hayward-like metric with the Schwarzschild prediction. Because the deviation is positive, which is different from that for the standard Hayward black hole, the regular-core scale of the standard Hayward black hole often leads to a decrease (negative deviation) in the deflection angle compared with a Schwarzschild black hole of the same mass. We conclude that the $\ell$ has slightly amplified the deflection. The positive deviation increases with the regular-core scale $\ell$. In addition, the deflection angle increases as the regular-core scale $\ell$ increases, as illustrated in Fig. \ref{fig:deflection_impact_ell}.

In the weak-field regime and under the small-angle approximation, the lens equation~\eqref{eq:lens_equation} simplifies to
\begin{equation}
\beta=\theta-\frac{D_{LS}}{D_S}\alpha\,,
\end{equation}
where $D_S=D_L+D_{LS}$. For perfect alignment ($\beta=0$), the image forms an Einstein ring at $\theta=\theta_E$. Using Eq.~\eqref{eq:deflection_angle} together with $b=D_L\theta_E$, we obtain the implicit equation
\begin{equation}
\theta_{E}
=\frac{D_{LS}}{D_{S}}
\left(
\frac{4m}{D_L\,\theta_E}
+\frac{15\pi}{4}\frac{m^{2}}{(D_L\,\theta_E)^{2}}
+\frac{16}{3}\frac{m\ell^{2}}{(D_L\,\theta_E)^{3}}
+\mathcal{O}\!\left(m^3,m^2\ell^2\right)
\right).
\label{eq:Einstein_ring}
\end{equation}
At the leading order, this simplifies to the familiar point-mass expression
\begin{equation}
\theta_E^{(0)}=\sqrt{\frac{4m\,D_{LS}}{D_L D_S}}\,,
\end{equation}
while the higher-order terms generate smaller positive corrections to $\theta_E$. In particular, since $\ell$ enters the deflection only at order $1/b^3$, its contribution to the Einstein ring is even weaker for astrophysical configurations where $b=D_L\theta_E\gg m$. In the Appendix \ref{sec:Einstein Ring of Galaxy}, we show that the calculated Einstein ring using observational data in ESO~325-G004 is consistent with the observed value of the Einstein ring in ESO~325-G004 ~\cite{Smith:2005pq,Smith:2013ena}. The theoretical results of deflection angles and Einstein rings are consistently slightly larger than those in the Schwarzschild case. This implies that while the current model is consistent with available data, future observations with higher angular resolution might be able to constrain $\ell$.

\section{Strong Field Limit and Observables}
\label{sec:SDL_observables}

\begin{figure}[t!]
\centering
    \includegraphics[width=0.55\textwidth]{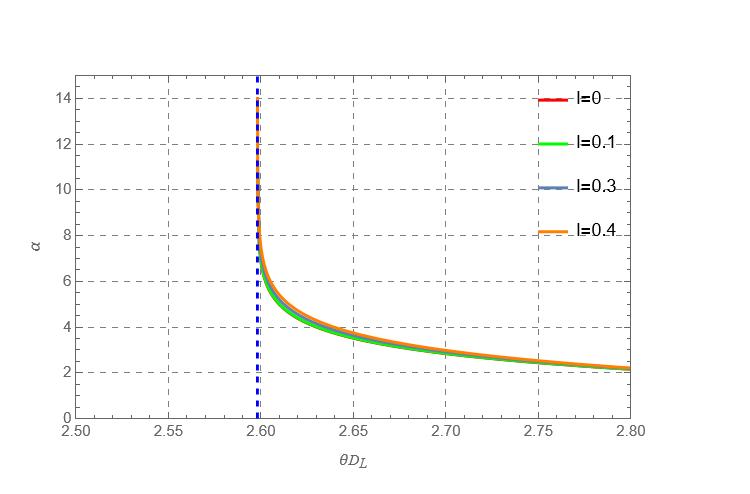}
	\caption{Deflection angle $\alpha$ (in radians) plotted against the impact parameter $b \approx \theta D_L$ (normalized to the Schwarzschild radius $R_s = 2m$) for various values of $\ell$. Vertical dashed line indicates the critical impact parameter $b_{ps} \approx 3\sqrt{3}m$, corresponding to the photon sphere, where the deflection angle diverges logarithmically.}
	\label{fig:SDL_deflection}
\end{figure}

While the corrections introduced by the regular-core scale $\ell$ are minimal in the weak-field regime, its signature is expected to be far more pronounced in the strong-gravity region near the black hole. The lensing in this regime is called strong field lensing. To investigate this, we employ the Strong Deflection Limit (SDL) method, which provides an analytic description of the relativistic images formed by photons orbiting near the photon sphere of the black hole~\cite{Bozza:2001xd,Bozza:2002zj}.

Identifying the observational fingerprints of the Hayward-like black hole requires a detailed analysis of its strong lensing properties. It is a well-established result that the deflection angle diverges logarithmically as the impact parameter approaches the critical value $b_{ps}$ corresponding to the photon sphere radius $\rho_{ps}$~\cite{Bozza:2001xd,Bozza:2002zj,Virbhadra:1999nm, Tsukamoto:2016jzh}. The radius of the photon sphere of the Hayward-like metric \eqref{eq:Hayward_like_metric} is defined as the largest root of the equation:
\begin{equation}
\frac{\left(r(\rho)\right)'}{r(\rho)}=\frac{\left(1-\frac{2M(\rho)}{\rho}\right)'}{\left(1-\frac{2M(\rho)}{\rho}\right)}\;,
\label{eq:photon_sphere_1}
\end{equation}
where the prime denotes differentiation with respect to $\rho$. For the Hayward-like metric under consideration, equation \eqref{eq:photon_sphere_1} simplifies to:
\begin{equation}
\rho_{ps}^3+2m\ell^2=3m\rho_{ps}^2\;.
\label{eq:photon_sphere_2}
\end{equation}
Around the photon sphere $b \to b_{ps}$, the deflection angle can be expanded as:
\begin{equation}
\alpha(b)=-\bar{a}\ln\left(\frac{b}{b_{ps}}-1\right)+\bar{b}+\mathcal{O}(b-b_{ps})\;,
\label{eq:SDL_deflection}
\end{equation}
where $\bar{a}$ and $\bar{b}$ are the SDL coefficients characterizing the deflection angle.

We first determine the critical impact parameter. By rewriting the metric \eqref{eq:Hayward_like_metric} components in terms of the generalized areal radius $r(\rho)$, the line element becomes:
\begin{equation}
ds^2=-\left(1-\frac{2m}{r(\rho)}\right)dt^{2}
+\left(1-\frac{2m}{r(\rho)}\right)^{-1}d\rho^{2}
+r^2(\rho)\,d\Omega^{2}\,.
\end{equation}
A remarkable feature of this specific Hayward-like construction is that the condition for the photon sphere, when expressed in terms of the areal radius $r(\rho)$, mirrors that of the Schwarzschild metric: $r(\rho_{ps})=3m$. Consequently, the critical impact parameter $b_{ps}$, which determines the angular size of the black hole shadow, reads:
\begin{equation}
b_{ps}=\sqrt{\frac{r^2(\rho_{ps})}{1-\frac{2m}{r(\rho_{ps})}}}=3\sqrt{3}m\;.
\end{equation}
It is independent of $\ell$ and retains its exact Schwarzschild value. The coefficient $\bar{a}$ is then analytically expressed as:
\begin{equation}
\bar{a}=\sqrt{\frac{2 B_{ps}A_{ps}}{(D\rho^2)''_{ps}B_{ps}-(D_{ps}\rho^2_{ps})B''_{ps}}}=\frac{\rho_{ps}}{3\rho_{ps}-6m}\;.
\end{equation}
In the limit $\ell \to 0$ (where $\rho_{ps} \to 3m$), we immediately recover the Schwarzschild result $\bar{a}=1$. Due to the analytical complexity of the coefficient $\bar b$, we evaluate the coefficient $\bar{b}$ numerically.

\begin{figure}[h!]
\centering
        \includegraphics[width=0.45\textwidth]{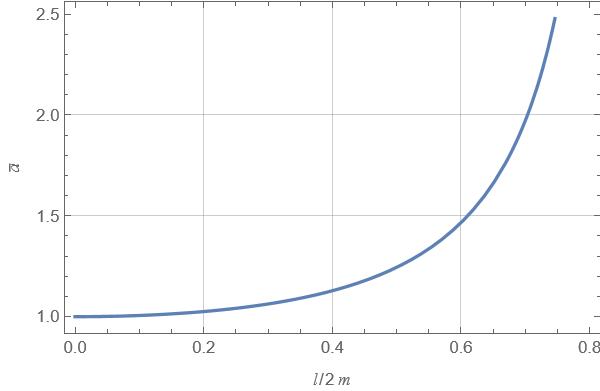}
        
	\caption{SDL coefficient $\bar{a}$ as a function of $\ell$: $\bar{a}$ increases with $\ell$.}
	\label{fig:abar_Haywardlike}
\end{figure}

\begin{figure}[h!]
\centering
   
        \includegraphics[width=0.45\textwidth]{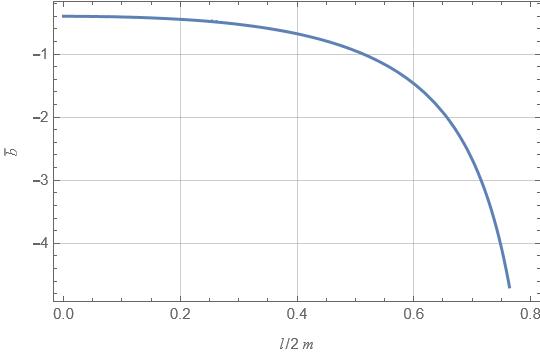}
        
	\caption{Behavior of the SDL coefficient $\bar b$ as a function of $\ell$: $\bar{b}$ decreases as $\ell$ increases.}
	\label{fig:bbar_Haywardlike}
\end{figure}

The behavior of $\bar a$ and $\bar b$ is shown in Fig.~\ref{fig:abar_Haywardlike}, Fig. \ref{fig:bbar_Haywardlike}, and Table~\ref{table:table_parameter}. We observe that $\bar{a}$ monotonically increases with $\ell$, whereas $\bar{b}$ decreases as $\ell$ increases. Our numerical results for $\ell=0$ recover the standard Schwarzschild values ($\bar{a}=1, \bar{b} \approx -0.40023$).

\begin{table}[h!]
    \centering
    \caption{Estimates for the SDL coefficients $\bar{a}$, $\bar{b}$, and the critical impact parameter $b_{ps}$ (in units of the Schwarzschild radius $R_s=2m$). Row $\ell=0$ corresponds to the Schwarzschild limit.}
    \label{table:table_parameter}
    \setlength{\tabcolsep}{10pt}
    \renewcommand{\arraystretch}{1.3}
    \begin{tabular}{l c c c}
        \hline\hline
        $\ell/2m$ & $\bar{a}$ & $\bar{b}$ & $b_{ps}/R_s$ \\
        \hline
        $0$    & 1.00000 & -0.40023  & 2.5981 \\
        $0.1$  & 1.00603 & -0.410826 & 2.5981 \\
        $0.3$  & 1.06296 & -0.528106 & 2.5981 \\
        $0.4$  & 1.12865 & -0.677135 & 2.5981 \\
        \hline\hline
    \end{tabular}
\end{table}

The resulting strong-field deflection angle is plotted in Fig.~\ref{fig:SDL_deflection}. As the impact parameter approaches $b_{ps}$, $\alpha$ exceeds $2\pi$, signifying that photons complete at least one loop around the black hole before escaping to the observer.

\subsection{Lens Equation and Relativistic Images}

The strong deflection regime, where the angles $\beta$ and $\theta$ are small, allows for the small-angle approximation in the lens equation~\cite{Bozza:2001xd,Bozza:2008ev,Virbhadra:1999nm}:
\begin{equation}
\beta = \theta - \frac{D_{LS}}{D_S}\Delta\alpha_n\;,
\label{eq:SDL_lens_eq}
\end{equation}
where $\Delta\alpha_n = \alpha(\theta) - 2n\pi$ is the effective deflection after subtracting $n$ full loops. Substituting the expansion in Eq.~\eqref{eq:SDL_deflection} into the lens equation \eqref{eq:SDL_lens_eq}, the position of the $n$-th relativistic image is given by \cite{Bozza:2002zj}:
\begin{equation}
\theta_{n} \simeq \theta_{n}^{0} + \frac{b_{ps}e_n(\beta-\theta_{n}^{0})D_{S}}{\bar{a}D_{LS}D_{L}}\;,
\label{eq:position_SDL_image}
\end{equation}
where $e_n = \exp\left(\frac{\bar{b}-2n\pi}{\bar{a}}\right)$, and $\theta_{n}^{0}$ represents the image position for a perfect $2n\pi$ deflection:
\begin{equation}
\theta_{n}^{0} = \frac{b_{ps}(1+e_n)}{D_L}\;.
\end{equation}
The magnification of these images, $\mu_n$, is determined by the conservation of surface brightness. Using Eq.~\eqref{eq:position_SDL_image} in the small-angle limit, the magnification for the $n$-th image is~\cite{Bozza:2002zj}:
\begin{equation}
\mu_n = \left.\left(\frac{\beta}{\theta}\frac{d\beta}{d\theta}\right)^{-1}\right|_{\theta^0_n} = \frac{b^2_{ps}e_n(1+e_n)D_S}{\bar{a}\beta D_{LS}D^2_{L}}\;.
\label{eq:SDL_magnification}
\end{equation}
The first relativistic image ($\theta_1$, $n=1$) is the brightest, as magnification decays exponentially with $n$ (since $e_n \ll 1$). For perfect alignment ($\beta=0$), these expressions diverge, indicating the formation of an Einstein ring.

To connect the theoretical formalism with potential astronomical observations, we consider three primary observables defined by Bozza~\cite{Bozza:2001xd}. The first is the asymptotic position of the set of images, $\theta_{\infty} = b_{ps}/D_L$. The second observable is the angular separation, $s = \theta_1 - \theta_{\infty}$, between the first relativistic image and the rest of the packed images. Physically, $s$ represents the visible "thickness" between the first relativistic image and the rest of the packed images; if $s$ is too small, the ring cannot be resolved from the rest of the images by current telescopes. The third observable is the relative flux ratios, $r_{\mathrm{mag}}$, defined as the ratio of the flux of the first image to the sum of the fluxes of all subsequent images (converted to a logarithmic magnitude scale). This metric quantifies the dominance of the outermost ring in the total luminosity. These are explicitly calculated as \cite{Bozza:2002zj}:
\begin{equation}
\begin{split}
    s &\approx \theta_{\infty} \exp{\left(\frac{\bar{b}-2\pi}{\bar{a}}\right)}\;,\\ 
r_{\mathrm{mag}} &= 2.5 \log_{10}\left(\frac{\mu_1}{\sum^{\infty}_{n=2}\mu_n}\right) \approx \frac{5\pi}{\bar{a}\ln{10}}\;.
\end{split}
\end{equation}
Since relativistic images arise from photons that encircle the black hole different 
number of times, another crucial and natural observable in the SDL is the time delays among these images. Following the standard treatment of 
strong-deflection lensing, we restrict our attention to the nearly aligned 
configuration, where the source angular offset is small ($\beta \approx 0$). 
The time delay between the $n$-th and $m$-th images located on the same side 
of the lens is given by \cite{Bozza:2003cp, Lu:2016gsf}:
\begin{equation}
\Delta T_{n,m} = 2\pi(n-m)b_{ps} + 2\sqrt{\frac{A_{ps}}{B_{ps}}}\sqrt{\frac{b_{ps}}{\hat{c}}} 
e^{\frac{\bar{b}}{2\bar{a}}} \left( e^{-\frac{m\pi}{\bar{a}}} - e^{-\frac{n\pi}{\bar{a}}} \right),
\label{eq:time_delay}
\end{equation}
where $A_{ps}$ and $B_{ps}$ are the matrix components in \eqref{eq:metric_lens} evaluated at $\rho_{ps}$ and $\hat{c}$ is a metric-dependent coefficient defined in \cite{Bozza_2002} as:
\begin{equation}
\hat{c} = \hat{\beta}_{ps} \sqrt{\frac{B_{ps}}{(D_{ps}\rho^2_{ps})^3}} \frac{\left((D\rho^2)'_{ps}\right)^2}{2(1-B_{ps})^2},
\end{equation}
where $'$ represents the derivative with respect to $\rho$.
\subsection{Observables for Sgr~A* and M87*}

We now estimate these lensing observables for the supermassive black hole candidates Sgr~A* and M87*, which currently serve as the premier astrophysical testbeds for strong-field gravity due to the unprecedented horizon-scale resolution achieved by the Event Horizon Telescope (EHT). Assuming these black hole candidates are Hayward-like, we adopt the following standard observational parameters: for M87*, a mass of $M=(6.5\pm 0.7)\times 10^9 M_\odot$ and $16.8$ Mpc~\cite{EventHorizonTelescope:2019pgp,EventHorizonTelescope:2019ggy}from us; and for Sgr~A*, a mass of $M=4.28^{+0.21}_{-0.1}\times 10^6 M_{\odot}$ and $8.32^{+0.07}_{-0.14}$ kpc from us, derived from stellar dynamics~\cite{gillessen2017update}.

The results of our calculations are presented in Tables~\ref{table:table_SDL_observables} and \ref{table:table_deviation_observables}. Notably, the asymptotic position of the relativistic images, $\theta_{\infty}$, remains constant ($19.80\,\mu\text{as}$ for M87* and $26.53\,\mu\text{as}$ for Sgr~A*) regardless of $\ell$. This implies a strong observational degeneracy: A measurement of $\theta_{\infty}$ alone cannot distinguish the Hayward-like black hole from a standard Schwarzschild black hole and impose no further constraint on the $\ell$. Consequently, breaking this degeneracy heavily relies on the measurement of other observables, which could manifest the effect of $\ell$. Nevertheless, $\theta_{\infty}$ can distinguish the Hayward-like black hole from the standard Hayward black hole, where the shadow size monotonically decreases with increasing $\ell$~\cite{Zhao:2017cwk}.

\begin{figure}[h!]
        \includegraphics[width=0.45\textwidth]{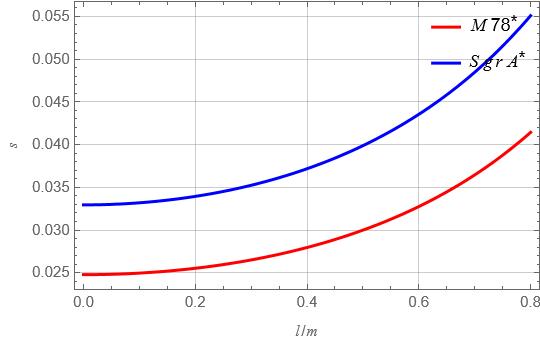}
    \caption{Angular separation $s$ for both Sgr~A* (blue) and M87* (red) increases with $\ell$.}
    \label{fig:s_SDL}
\end{figure}

\begin{figure}[h!]
\centering
        \includegraphics[width=0.45\textwidth]{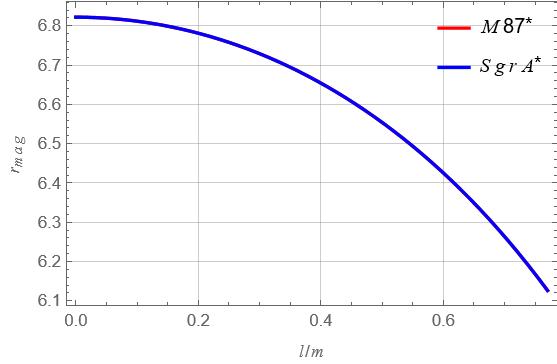}
        
    \caption{Relative flux ratio $r_{\mathrm{mag}}$ for both Sgr~A* (blue) and M87* (red) decreases as $\ell$ increases. Note that $r_{\mathrm{mag}}$ is independent of the mass-to-distance ratio, resulting in overlapping curves for both sources.}
    \label{fig:rm_SDL}
\end{figure}

Conversely, the second lensing observables exhibit clear, $\ell$-dependent deviations that could possibly be probed by future-generation interferometry. The angular separation $s$ increases with $\ell$ (see Fig.~\ref{fig:s_SDL}), meaning that the first relativistic image appears slightly further from the rest of images compared with the Schwarzschild case and has value ranging from 33 to 53 nanoarcsecond (nas) for Sgr~A* and 25 to 40 nanoarcsecond (nas) for M87*, where $\ell$ is from $0$ to $0.77m$ for both cases. Meanwhile, the relative flux ratio $r_{\mathrm{mag}}$ decreases from 6.82 to 6.13 as $\ell$ increases (Fig.~\ref{fig:rm_SDL}). Both of these observables share the same qualitative behavior as those in the standard Hayward case~\cite{Zhao:2017cwk}. Although the relative flux ratio $r_{\mathrm{mag}}$ serves as our main observational proxy, the specific magnifications ($\mu_n$) and angular positions ($\theta_n$) of the individual relativistic images remain strongly sensitive to the source alignment $\beta$. For the Hayward-like black hole, Tables~\ref{table:table_magnification_Sgr} and \ref{table:table_magnification_M87} record, respectively, for Sgr~A* and M87* the numerical values of $\mu_n$ and $\theta_n$ for both primary and secondary images. As evident in these results, the higher-order relativistic images ($\theta_n$ for $n>1$) pack tightly together, asymptotically approaching the shadow $\theta_{\infty}$. 

\begin{figure}[h!]
\centering
        \includegraphics[width=0.45\textwidth]{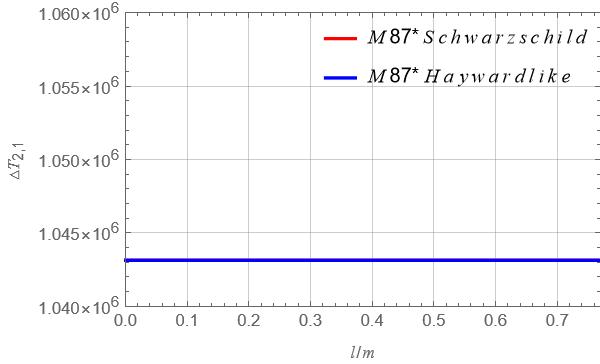}
        
	\caption{Total time delay between the first and second relativistic images for M87* as a function of $\ell$. Detailed changing behavior is shown in Fig.~\ref{fig:timedelay_SDL_M87_second}.}
	\label{fig:timedelay_SDL_M87}
\end{figure}

\begin{figure}[h!]
\centering
        \includegraphics[width=0.45\textwidth]{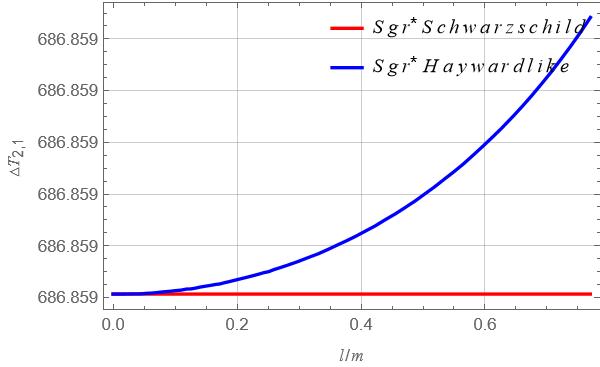}
	\caption{Total Time delay between the first and second relativistic images and varies with $\ell$  for Sgr~A*. The time delay increases with $\ell$.}
	\label{fig:timedelay_SDL_sgr}
\end{figure}

\begin{figure}[h!]
\centering
        \includegraphics[width=0.45\textwidth]{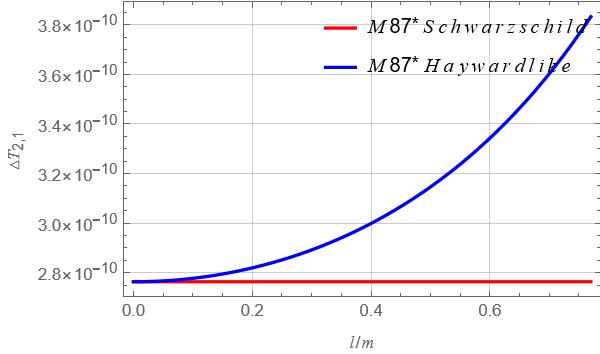}
        
	\caption{Time delay deviation from Schwarzschild between the first and second relativistic images for M87* as a function of $\ell$ (in seconds). The deviation increases as $\ell$ increases.}
	\label{fig:timedelay_SDL_M87_second}
\end{figure}

\begin{figure}[h!]
\centering
        \includegraphics[width=0.45\textwidth]{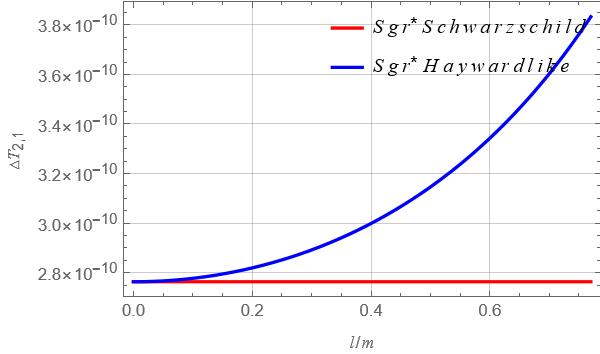}
	\caption{Time delay deviation from Schwarzschild between the first and second relativistic images as a function of $\ell$ for Sgr A*.}
	\label{fig:timedelay_SDL_sgr_second}
\end{figure}

Turning to the temporal observables, the total time delay $\Delta T_{2,1}$ between the first and second images is plotted in Figs.~\ref{fig:timedelay_SDL_M87} and \ref{fig:timedelay_SDL_sgr} using the respective mass and distance estimates for M87* and Sgr~A*. For both cases, time delay increases with $\ell$. Because the time scale ($m/c$) for M87* is so large that the $\ell$-dependent time delay is imperceptible on the plot of total time delay. This increasing behavior and the deviation from the standard Schwarzschild prediction can be better observed in Figs.~\ref {fig:timedelay_SDL_M87_second} and \ref{fig:timedelay_SDL_sgr_second}, where the growth in $\Delta T_{2,1}$ is driven specifically by the second term in Eq.~\eqref{eq:time_delay}.

\begin{table}[ht]
    \centering
    \setlength{\tabcolsep}{5pt}
    \renewcommand{\arraystretch}{1.3}
    
    \caption{Lensing observables for M87* and Sgr~A*. Row $\ell=0$ represents the standard Schwarzschild reference values.}
    \label{table:table_SDL_observables}
    
    \begin{tabular}{l c c c c c}
        \hline\hline
         & \multicolumn{2}{c}{\textbf{M87*}} & \multicolumn{2}{c}{\textbf{Sgr A*}} & \\
        $\ell/m$ & $\theta_{\infty} (\mu\text{as})$ & $s (\mu\text{as})$ & $\theta_{\infty} (\mu\text{as})$ & $s (\mu\text{as})$ & $r_{\mathrm{mag}}$ \\
        \hline
        0.0  & 19.8012 & 0.02748 & 26.3274 & 0.03295 & 6.822 \\
        0.1  & 19.8012 & 0.02497 & 26.3274 & 0.03319 & 6.812 \\
        0.3  & 19.8012 & 0.02651 & 26.3274 & 0.03524 & 6.729 \\
        0.5  & 19.8012 & 0.03000 & 26.3274 & 0.03988 & 6.554 \\
        0.77 & 19.8012 & 0.03974 & 26.3274 & 0.05289 & 6.126 \\
        \hline\hline
    \end{tabular}
    
    \vspace{0.5cm}
    
    \caption{Deviations from the Schwarzschild limit ($\ell=0$) for the relevant lensing observables.}
    \label{table:table_deviation_observables}
    
    \begin{tabular}{l c c c c c}
        \hline\hline
         & \multicolumn{2}{c}{\textbf{M87*}} & \multicolumn{2}{c}{\textbf{Sgr A*}} & \\
        $\ell/m$ & $\delta\theta_{\infty} (\mu\text{as})$ & $\delta s (\mu\text{as})$ & $\delta\theta_{\infty} (\mu\text{as})$ & $\delta s (\mu\text{as})$ & $\delta r_{\mathrm{mag}}$ \\
        \hline
        0.0  & 0 & 0           & 0 & 0           & 0 \\
        0.1  & 0 & $+0.00018$  & 0 & $+0.00025$  & $-0.010$ \\
        0.3  & 0 & $+0.00173$  & 0 & $+0.00230$  & $-0.093$ \\
        0.5  & 0 & $+0.00521$  & 0 & $+0.00693$  & $-0.268$ \\
        0.77 & 0 & $+0.01500$  & 0 & $+0.01995$  & $-0.696$ \\
        \hline\hline
    \end{tabular}
\end{table}
\begin{table*}[ht]
    \centering
    \setlength{\tabcolsep}{4pt}
    \renewcommand{\arraystretch}{1.0}
    
\caption{Magnifications and image positions of the first two relativistic images due to lensing by Sgr A* for $\beta=1$ with $D_{LS}/D_S=1/2$.}
    \label{table:table_magnification_Sgr}
    
    \begin{tabular}{l c c c c c c c c}
        \hline\hline
         & \multicolumn{8}{c}{\textbf{Sgr A*}}  \\
         $\ell/m$ & $\mu_{p,1}(10^{-12})$ & $\mu_{p,2}(10^{-14})$ & $\mu_{s,1}(10^{-12})$  & $\mu_{s,2}(10^{-14})$ & $\theta_{p,1} (\mu\text{as})$ & $\theta_{p,2} (\mu\text{as})$ & $\theta_{s,1} (\mu\text{as})$ & $\theta_{s,2} (\mu\text{as})$ \\
        \hline
         0.0  & 8.43  & 1.57  & $-8.43$  & $-1.57$  & 26.3603 & 26.3274 & 26.3603 & 26.3274 \\
         0.1  & 8.49  & 1.60  & $-8.49$  & $-1.60$  & 26.3605 & 26.3274 & 26.3605 & 26.3274 \\
         0.3  & 8.90  & 1.81  & $-8.90$  & $-1.81$  & 26.3626 & 26.3274 & 26.3626 & 26.3274 \\
         0.5  & 9.81  & 2.35  & $-9.81$  & $-2.35$  & 26.3672 & 26.3274 & 26.3672 & 26.3274 \\
         0.77 & 12.20 & 4.31  & $-12.20$ & $-4.31$  & 26.3802 & 26.3275 & 26.3802 & 26.3275 \\
        \hline\hline
    \end{tabular}

    \vspace{0.8cm}
    
    \caption{Magnifications and image positions of the first two relativistic images due to lensing by M87* for $\beta=1$ with $D_{LS}/D_S=1/2$.}
    \label{table:table_magnification_M87}
    
    \begin{tabular}{l c c c c c c c c}
        \hline\hline
         & \multicolumn{8}{c}{\textbf{M87* }}  \\
         $\ell/m$ & $\mu_{p,1}(10^{-12})$ & $\mu_{p,2}(10^{-15})$ & $\mu_{s,1}(10^{-12})$  & $\mu_{s,2}(10^{-15})$ & $\theta_{p,1} (\mu\text{as})$ & $\theta_{p,2} (\mu\text{as})$ & $\theta_{s,1} (\mu\text{as})$ & $\theta_{s,2} (\mu\text{as})$ \\
        \hline
         0.0  & 4.77 & 8.90  & $-4.77$ & $-8.90$  & 19.8259 & 19.8012 & 19.8259 & 19.8012 \\
         0.1  & 4.80 & 9.03  & $-4.80$ & $-9.03$  & 19.8261 & 19.8012 & 19.8261 & 19.8012 \\
         0.3  & 5.03 & 10.20 & $-5.03$ & $-10.20$ & 19.8277 & 19.8012 & 19.8277 & 19.8012 \\
         0.5  & 5.55 & 13.20 & $-5.55$ & $-13.20$ & 19.8312 & 19.8012 & 19.8312 & 19.8012 \\
         0.77 & 6.88 & 24.30 & $-6.88$ & $-24.30$ & 19.8409 & 19.8013 & 19.8409 & 19.8013 \\
        \hline\hline
    \end{tabular}
    
\end{table*}
To validate our results by current observations, we first consider the angular diameter of the shadow of the M87*. The EHT collaboration reported a measured emission ring diameter of $\theta_{sh}=42\pm3\,\mu\text{as}$~\cite{EventHorizonTelescope:2019dse,EventHorizonTelescope:2019pgp,EventHorizonTelescope:2019ggy}. We use the apparent radius of the photon sphere, $\theta_{\infty}$, as a theoretical proxy for the angular size of the black hole shadow. As shown in Table~\ref{table:table_SDL_observables}, $\theta_{\infty}=19.8012\,\mu\text{as}$ and remains $\ell$-independent. The resulting shadow diameter is $\theta_{\mathrm{sh}}=2\theta_{\infty}=39.6024\,\mu\text{as}$, falling within the error bars of the EHT data.

Similarly, in 2022, the EHT targeted at the Sgr~A* at the our Milky Way's center, revealing a shadow image with an extracted diameter of $\theta_{\mathrm{sh}}=48.7\pm 7\,\mu\text{as}$~\cite{EventHorizonTelescope:2022wkp}. This observation suggest a black hole mass of $M=4.0^{+1.1}_{-0.6}\times 10^6 M_{\odot}$ and a Schwarzschild shadow deviation of $\delta=-0.08^{+0.09}_{-0.09}$ (when calibrated with VLTI priors) or $-0.04^{+0.09}_{-0.10}$ (with Keck priors). Such findings provide a new methodology to probe the strong-field regime~\cite{EventHorizonTelescope:2022apq,EventHorizonTelescope:2022wkp, EventHorizonTelescope:2022urf}. The EHT collaboration utilized three independent algorithms to determine that the mean measured value of the shadow angular diameter $\theta_{\mathrm{sh}}$ falls within the range $(46.9,50.0)\,\mu\text{as}$ and within $(41.7,55.7)\,\mu\text{as}$ for $1\sigma$ interval ~\cite{EventHorizonTelescope:2022xqj,EventHorizonTelescope:2022wkp}. From Table~\ref{table:table_SDL_observables}, our calculated value for Sgr~A* is $\theta_{\infty}=26.3274\,\mu\text{as}$, which yields a total shadow diameter for the Hayward-like black hole of $2\theta_{\infty}=52.6548\,\mu\text{as}$. This theoretical prediction falls well within the EHT's observed interval and the error bars, justifying the viability of the Hayward-like metric as a phenomenological alternative to Schwarzschild. To constrain the parameters $\ell$ of the Hayward-like black hole, however, requires either separation $s$ or the relative flux ratio $r_{\mathrm{mag}}$.

\section{Conclusion and discussion}
\label{sec:conclusion}

In this work, we began by examining gravitational lensing in the weak-field regime for the Hayward-like spacetime. We observed that the deflection angle exhibits a positive deviation, increasing with the parameter $\ell$. This behavior provides a distinct contrast to the standard Hayward black hole, which typically yields a negative deviation. To justify our prediction fit within current astrophysical bounds, we applied it to the observed Einstein ring ($\theta_{E}$) of the ESO~325-G004. Assuming the total lens mass consists of a central Hayward-like black hole embedded within the galaxy's visible and dark matter, our theoretically derived $\theta_{E}$ remains strictly consistent with the $1\sigma$ observational uncertainty window of $2.85 \pm 0.40$ arcsec.

Transitioning to the strong-field regime, we utilized the SDL framework. By analytically extracting the coefficient $\bar{a}$ and numerically evaluating $\bar{b}$, we determined the strong-field deflection angle along with lensing observables: the asymptotic position $\theta_{\infty}$, the angular separation $s$, and the relative flux ratio $r_{\mathrm{mag}}$. Our analysis shows that for a fixed impact parameter, the strong-field deflection angle exceeds that of the Schwarzschild metric and grows monotonically as $\ell$ increases.

To evaluate the prospective observability of these strong-field phenomena, we modeled Sgr~A* and M87* as Hayward-like black holes to estimate their theoretical SDL parameters and observables. While current EHT limits cannot yet probe these precise geometric scales, future interferometric facilities aiming for angular resolutions near $100\,\text{nas}$ (nanoarcseconds) could potentially differentiate the Hayward-like black hole from the standard Hayward black hole based solely on accurate measurements of the asymptotic position $\theta_{\infty}$. 

If observational resolutions eventually approach $\sim 10\,\text{nas}$ or better, visually separating the first relativistic image from the photon ring may be feasible. It is possible to measure observables $s$ and $r_{\mathrm{mag}}$, potentially constraining the $\ell$ in the Hayward-like black hole and distinguishing it from the classical Schwarzschild black hole. Because regular black holes can potentially be related to exact vacuum solutions of generally covariant gravity theories~\cite{Zhang:2025ccx}, ultra-high-resolution lensing presents a compelling, albeit distant, frontier for testing fundamental gravity. We, however, emphasize that resolving such scales remains well beyond the capacity of contemporary technology. Consequently, alternative phenomenological avenues -- such as the enhanced magnifications predicted in the microlensing of regular black holes~\cite{Boos:2025nzc} -- may serve as more accessible observational targets in the near future. Investigating the specific microlensing signatures of Hayward-like spacetimes in this context remains a highly relevant direction for upcoming research.

Beyond the standard SDL parameters, we also provided an exhaustive tabulation of the signed magnifications and angular positions for the first and second relativistic images, alongside the differential time delay $\Delta T_{2,1}$ between them. Consequently, these features can be broadly categorized into ``imaging'' observables (which are severely constrained by current angular resolution and dynamic range limitations) and ``timing'' observables like $\Delta T_{2,1}$. In the geometric-optics approximation, if the background source happens to be a compact, time-variable radio emitter, $\Delta T_{2,1}$ might theoretically be detected as a phase lag between two highly demagnified signal echoes.

While this framework lays a phenomenological foundation for future strong-gravity radio-timing studies -- perhaps eventually aligning with next-generation arrays like the SKA -- we must maintain a grounded perspective. In the Hayward-like black hole presented here, the predicted deviation in $\Delta T_{2,1}$ from the Schwarzschild baseline is exceedingly small. Therefore, its role at the current stage is primarily conceptual rather than a readily accessible diagnostic tool. Nonetheless, as both weak- and strong-field observational techniques advance, gravitational lensing remains a potential future opportunity for probing Hayward-like and other geometries and the fundamental nature of spacetime.

\begin{acknowledgements}
The authors thank Liang Dai for his introducing and inviting us to the area of gravitational lensing and for his insightful comments on our work. The authors also thank Huanyuan Shan, Long Wang, Le Zhang, Yi Zheng, and Zhiqi Huang for their helpful comments. YW is supported by NSFC Grant No. 12475001, the Shanghai Municipal Science and Technology Major Project (Grant No. 2019SHZDZX01), Science and Technology Commission of Shanghai Municipality (Grant No. 24LZ1400100), and the Innovation Program for Quantum Science and Technology (No. 2024ZD0300101). YW is grateful for the hospitality of the Perimeter Institute during his visit, where this work was partially done. This research was supported in part by the Perimeter Institute for Theoretical Physics. Research at Perimeter Institute is supported by the Government of Canada through the Department of Innovation, Science and Economic Development and by the Province of Ontario through the Ministry of Research, Innovation and Science.
\end{acknowledgements}


\raggedbottom
\appendix
\nolinenumbers

\section{GBT method}
\label{appendix:GBT}
The strategy of the GBT method is as follows. The Gauss--Bonnet theorem for a domain $D$ reads \begin{equation}
\iint_{D} K\, dS+\int_{\partial D}\kappa\, dt+\sum_i \phi_i
=2\pi\,\chi(D)\,,
\label{eq:GBT_formula}
\end{equation}
where $t$ is an affine parameter along $\partial D$, $K$ is the Gaussian curvature of the optical metric, $\kappa$ is the geodesic curvature of $\partial D$, $\phi_i$ are the exterior angles at corner points, and $\chi(D)$ is the Euler characteristic of $D$. The exterior angles are related to the corresponding interior angles by
\begin{equation}
\theta_S=\pi-\phi_S,\qquad \theta_O=\pi-\phi_O\,.
\end{equation}
We assume that the lens spacetime is non-singular in the relevant region so that $\chi(D)=1$. For a boundary segment that is a geodesic, the geodesic curvature vanishes by definition, $\kappa=0$. For the domain $D$ used in the standard GBT lensing setup shown in Fig. \ref{fig:GBT_graph}, the source $S$ and the observer $O$ are taken to lie in an asymptotically flat region, which implies $\theta_S=\theta_O=\pi/2$. In this case Eq.~\eqref{eq:GBT_formula} reduces to
\begin{equation}
\int_{\gamma_P}\kappa(\gamma_P)\,dt
=\pi-\iint_{D}K\,dS\,,
\end{equation}
where $\gamma_p$ denotes the large circular arc at infinity closing the domain. For an asymptotically flat optical metric one has $\kappa\,\frac{dt}{d\varphi}\to 1$ on $\gamma_p$, and therefore the deflection angle can be obtained directly from the Gaussian curvature as
\begin{equation}
\alpha=-\iint_{D}K\,dS\,.
\label{eq:deflection_GBT}
\end{equation}
In our case (see Eq.~\eqref{eq:Hayward_like_metric} with $\theta=\pi/2$), the optical line element on the equatorial plane $(\rho,\varphi)$ can be written in orthogonal coordinates as
\begin{equation}
dt^{2}=E(\rho)\,d\rho^{2}+G(\rho)\,d\varphi^{2}\,,
\end{equation}
so that the Gaussian curvature is
\begin{equation}
\begin{split}
K&=-\frac{1}{2\sqrt{EG}}\partial_{\rho}\!\left(\frac{\partial_{\rho}G}{\sqrt{EG}}\right)\\
&=-\frac{1}{2EG}\,\partial_{\rho}^{2}G
+\frac{1}{4}\frac{(\partial_{\rho}G)^{2}}{EG^{2}}
+\frac{1}{4}\frac{(\partial_{\rho}G)(\partial_{\rho}E)}{E^{2}G}\,.
\end{split}
\end{equation}

\section{Hayward-like BH}
\label{sec:topological_BH}
A Hayward-like black hole has the metric
\begin{equation}
ds^2=-\left(1-\frac{2M(\rho)}{\rho}\right)dt^{2}
+\left(1-\frac{2M(\rho)}{\rho}\right)^{-1}d\rho^{2}
+r^2(\rho)\,d\Omega^{2}\,,
\label{eq:Hayward_like_metric_1}
\end{equation}
where
\begin{equation}
M(\rho)=\frac{m\,\rho^{3}}{\rho^{3}+2m\ell^{2}},
\qquad
r(\rho)
=\rho+\frac{2m\ell^{2}}{\rho^{2}}\,.
\end{equation}
As $\rho\to\infty$, one has
\begin{equation}
1-\frac{2M(\rho)}{\rho}\to 1-\frac{2m}{\rho},
\qquad
r(\rho)\to \rho,
\end{equation}
so the geometry is an asymptotically Schwarzschild region. In the opposite limit $\rho\to 0^{+}$,
\begin{equation}
1-\frac{2M(\rho)}{\rho}\to 1-\frac{\rho^{2}}{\ell^{2}},
\qquad
r(\rho)\sim \frac{2m\ell^{2}}{\rho^{2}}\to \infty.
\end{equation}
Using $r(\rho)\sim 2m\ell^{2}/\rho^{2}$ for small $\rho$, we obtain
\begin{equation}
1-\frac{\rho^{2}}{\ell^{2}}
\sim 1-\frac{2m}{r}\,,
\end{equation}
so the asymptotic region for $\rho\to 0^+$ is also Schwarzschild (in terms of the areal radius $r$).

The function $r(\rho)$ attains its minimum $r_0=r(\rho_{0})=\frac{3}{2^{1/3}}(m\ell^{2})^{1/3} $, where $\rho_{0}=(4m\ell^{2})^{1/3}$.

Following~\cite{Calza:2025mrt}, one chooses the branch $\rho^{3}\ge 4m\ell^{2}$ so that $r(\rho)$ increases monotonically with $\rho$.

Two surfaces are of particular interest. The first is determined by the (outer) horizon condition,
\begin{equation}
1-\frac{2M(\rho_H)}{\rho_H}=0
\quad\Longleftrightarrow\quad
\rho_H^{3}+2m\ell^{2}=2m\rho_H^{2}.
\label{eq:horizon_rho}
\end{equation}
On the chosen branch, one finds $r(\rho_H)=2m$. Requiring the horizon to lie outside the minimal sphere, $r_H>r_0$, yields the bound
\begin{equation}
\ell<\frac{4m}{3\sqrt{3}}
\qquad
\text{equivalently}\qquad
4m>3\sqrt{3}\,\ell,
\end{equation}
in agreement with~\cite{Calza:2025mrt}. The second surface is the branch point
\begin{equation}
\rho_\ell^{3}=4m\ell^{2}.
\end{equation}
Since the branch of $\rho$ is restricted to $\rho^{3}\ge 4m\ell^{2}$, this restriction effectively replaces an inner (Cauchy) horizon as the inner boundary of the spacetime described by~\eqref{eq:Hayward_like_metric_1}.
\section{The Einstein Ring of Galaxy ESO 325-G004}
\label{sec:Einstein Ring of Galaxy}
We analyze how $\ell$ affects the Einstein-ring angular radius. We consider perfect alignment of the source, lens, and observer ($\beta=0$), with both the source and the observer located in nearly flat regions. We use observational data from the Einstein ring of the galaxy ESO~325-G004, whose inferred total lensing mass is $M=1.50\times 10^{11}M_{\odot}$~\cite{Smith:2005pq,Smith:2013ena}, where $M_{\odot}=1.98\times 10^{30}\,\mathrm{kg}$. This mass includes contributions from the central black hole and the galaxy's dark and luminous matter. The background source galaxy lies at a higher redshift, $z_s=2.141$.

\begin{figure}[h!]
\centering

\includegraphics[width=0.45\textwidth]{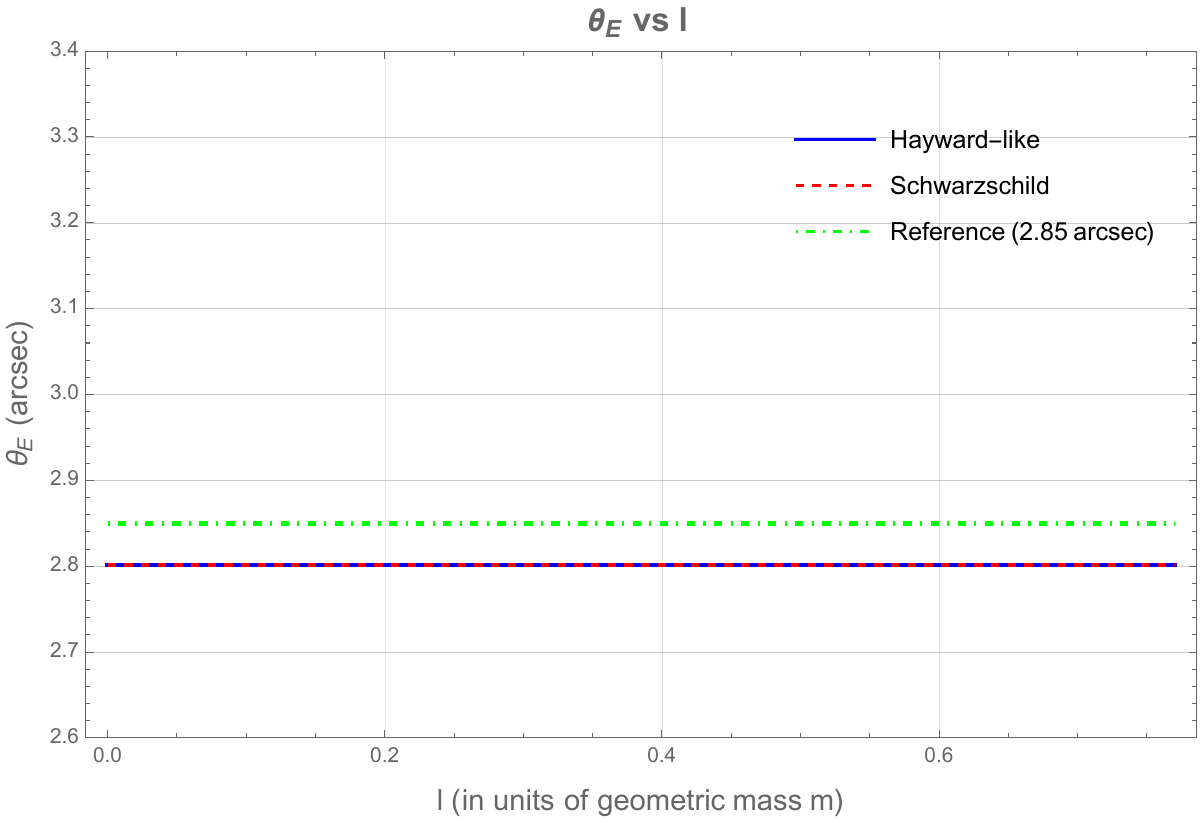}

	\caption{ Blue and red dotted lines represent the Einstein rings of Hayward-like BHs and Schwarzschild BHs, respectively, and the green dotted line represents the observed value of the Einstein ring $2.85$ arcsec and the range of the y-axis denoting the $1\sigma$ uncertainty. Results show that the Hayward-like lens reproduces the observed Einstein-ring size within the uncertainties over the $\ell$ range considered. } 
	\label{fig:theta_E}
\end{figure}

For a rough distance estimate, we adopt Hubble's law
\begin{equation}
cz=H_0 D\,,
\end{equation}
where $H_0=70.4\,\mathrm{km}\,\mathrm{s^{-1}}\,\mathrm{Mpc^{-1}}$ is the Hubble constant and $D$ is the proper distance. In a flat universe, we approximate the comoving distance by $d=(1+z)D$, such that the lens -- source and observer -- lens distances are
\begin{equation}
\begin{split}
D_{LS}&=\frac{cz_s(1+z_s)}{H_0}=2.866\times 10^4 \,\mathrm{Mpc}\;,\\
D_{L}&=\frac{cz_l(1+z_l)}{H_0}=1.544\times 10^2 \,\mathrm{Mpc}\;.
\end{split}
\end{equation}
The measured Einstein-ring radius of ESO~325-G004 at lens redshift $z_l=0.035$ is~\cite{Smith:2005pq,Smith:2013ena}
\begin{equation}
\theta_E=(2.85^{+0.55}_{-0.25})^{\prime\prime}\,.
\end{equation}
We solve the equation \eqref{eq:Einstein_ring} numerically using observational data in the ESO~325-G004 ~\cite{Smith:2005pq,Smith:2013ena} and present $\theta_E(\ell)$ in Fig.~\ref{fig:theta_E}. The result shows that the Hayward-like lens reproduces the observed Einstein ring size within the uncertainties over the $\ell$ range considered.  It also runs nearly parallel to the Schwarzschild prediction (red dotted line). This implies that galaxy-scale Einstein rings probe the weak-field region, where the leading Schwarzschild term dominates, and regular-core corrections are strongly suppressed.

\bibliographystyle{aa} 
\bibliography{QI}

\end{document}